\documentclass[12pt]{article}
\usepackage{amssymb,amsmath,amsfonts,amsthm,graphicx,epsfig}

\newcommand{\forget}[1]{\iffalse#1\fi}
\newcommand{\forgetmenot}[1]{\iftrue#1\fi}

\newcommand{\be}{\begin{equation}}
\newcommand{\ee}{\end{equation}}
\newcommand{\ba}{\begin{eqnarray}}
\newcommand{\ea}{\end{eqnarray}}

\newcommand{\lapp}{\mathrel{\vcenter{\hbox{\tiny \ooalign{\raise 3.25pt
        \hbox{$<$}\crcr $\sim$}}}}}
\newcommand{\gapp}{\mathrel{\vcenter{\hbox{\tiny \ooalign{\raise 3.25pt
        \hbox{$>$}\crcr $\sim$}}}}}
\newcommand{\eqdef}{\!\!\mathrel{\vcenter{\hbox{ \ooalign{\raise 4.75pt
        \hbox{${\textsf{\tiny{def}}}$}\crcr $=$}}}}}

\newcommand{\etal}{{\it{et~al.}}}

\begin{document}

\title{Cosmic microwave background limits on spatially homogeneous
cosmological models with a cosmological constant.}

\author{A. A. Coley and W. C. Lim}

\maketitle

{Department of Mathematics and Statistics, Dalhousie University, 
Halifax, Nova Scotia, Canada, B3H~3J5}

\begin{abstract}

We investigate the effect of dark energy on the limits on the
shear anisotropy in spatially homogeneous Bianchi cosmological
models obtained from measurements of the temperature anisotropies
in the cosmic microwave background. We shall primarily assume that
the dark energy is modelled by a cosmological constant. In
general, we find that there are tighter bounds on the shear than
in models with no cosmological constant, although the limits are
(Bianchi) model dependent. In addition, there are special
spatially homogeneous cosmological models whose rate of expansion
is highly anisotropic, but whose cosmic microwave background
temperature is measured to be exactly isotropic at one instant of
time.

\end{abstract}

\newpage

\section{Introduction}

Observations  suggest that the Universe has $0.98\leq
\Omega_{tot}\leq 1.08$  with radiation (R), baryons (B), dark
matter (DM) and dark energy (DE) each contributing $\Omega_R\simeq
5\times 10^{-5},\Omega_B\simeq 0.04,\Omega_{DM}\simeq
0.26,\Omega_{ DE}\simeq 0.7,$ respectively \cite{cmbr}, where the
variables $\Omega_i=\rho_i/ \rho_c$ give the fractional
contribution of different components of the Universe ($ i$
denoting baryons, dark matter, radiation, etc.) to the critical
energy density $\rho_c=3H^2_0/8\pi G$, where $H_0=(\dot a/a)_0$ is
the present Hubble expansion rate. Consequently, we have that
$\Omega_{0} \equiv   \Omega_R + \Omega_B + \Omega_{DM} \sim 0.3$.
The simplest model for a fluid with negative pressure is the
cosmological constant \cite{padmanabhan:2003} whence nearly
seventy per cent of the energy density in the Universe is
unclustered and exerts negative pressure; $\Omega_{\Lambda_0}
\equiv \Omega_{ DE} \sim 0.7$.

It is of interest to study the effect of dark energy on the
cosmological limits from cosmic microwave background (CMB)
measurements on the shear (and the rotation, etc.),
especially in spatially homogeneous (SH) cosmological models. In
the presence of the cosmological constant, an SH universe that is very
nearly flat now and that was very nearly flat at the time of last
scattering can become significantly negatively curved at
intermediate periods.  At earlier times the matter dominates and
the curvature is dynamically negligible, while at later times
$\Omega_\Lambda$ can dominate. For a currently almost flat
universe with parameters close to the observed values of
$\Omega_{\Lambda_0}\sim 0.7$ and $\Omega_{0}\sim 0.3$, the
curvature is at most only a few percent. At intermediate times,
however, the curvature can be dynamically significant.

The observed temperature anisotropy (from COBE  and WMAP \cite{bennett:2003})
of the CMB radiation, $\Delta
T/T$, on large angular scales is determined by the value of
the shear at the surface of last scattering of the radiation.
To evaluate large-scale anisotropies  we must evaluate the
peculiar redshift a photon will feel from the epoch of last
scattering (${\it ls}$) until now $(0)$ by integration along null
geodesics in the background spacetime.

The usual paradigm invoked to understand cosmological observations,
motivated by inflation and consistent with the smoothness 
of the CMB, is that the Universe was spatially homogeneous and
isotropic spacetime at early times. 
The theory is then that
some physical mechanism (such as inflation) generated
perturbations that subsequently evolved through gravitational collapse to form
the structures we now observe. Within this paradigm, the
Universe is closely modelled by a Friedmann-Robertson-Walker (FRW)
cosmology at all times since the earliest epoch (and particularly since
recombination). In such models it is often the 
case that the integrated effect from a small shear is not important.

We utilize dynamical systems methods using expansion-normalized
variables in SH cosmological models \cite{WE}, generalized to
include a cosmological constant, to revisit the cosmological
constraints on the Hubble-normalized shear scalar $\Sigma =
\sigma/(\sqrt{3}H)$ (for example) from CMB anisotropy measurements.

\section{Background}

Bianchi models provide a description of general SH anisotropic
cosmologies. There are distinct features depending on the overall
geometry and homogeneity class of the model \cite{Bianchi}. In a
pioneering paper,  Collins and Hawking used analytical arguments
to find upper bounds on the amount of shear (and
vorticity) in the Universe today, from the absence of
any detected CMB anisotropy \cite{shear}. A detailed numerical
analysis of such models \cite{BJS} used experimental limits on the
dipole and quadrupole to refine limits on the universal shear (and
rotation). More recently, it  has been argued \cite{Barrow} that
there is no `isotropy problem' in Bianchi type VII$_h$
cosmological models in which shear (and vorticity) have decayed
only logarithmically since the Planck time, in that the present
amplitude of CMB fluctuations is compatible with current
observational limits in these models.

The limits could be strengthened in a number of ways; for example,
if we assume particular classes of cosmological models (i.e.,
spatially homogeneous models) for which exact evolution laws are
known we can replace assumptions and approximations by exact
relations. For particular Bianchi models with specific evolution
equations, stronger limits are indeed possible. In particular, in
general shear anisotropy grows with time (i.e., Bianchi
models do not generically isotropize). However, for very special
models (i.e., not generic) of Bianchi type I and V the anisotropy
decays, and hence \emph{much} stronger limits on present day shear
may be obtained in these models.

In particular, in \cite{MGS} it was shown that small quadrupole
anisotropies in the CMB imply severe limits on spacetime
anisotropy in a Bianchi type I model (assumed to be a small
perturbation of a FRW model). Using an exact Bianchi I
cosmological solution, the  quadrupole component of CMB
temperature distribution found by  
COBE \cite{bennett:2003} (expressed as an upper limit on $\Delta
T /T$), implies that $\Sigma_0 \leq 2 \times 10{^{-9}}$. Similarly,
strong limits can be obtained in Bianchi type V models (and a
special class of spatially homogeneous Bianchi type VII$_h$ models
\cite{Barrow}). However, such severe limits are not generic.

The above limits on the shear (and rotation) are not based on
`full sky' maps; only the simple quadrupole (and not higher
multipoles, which produce hotspots/spirals) is utilized. In
general SH models the distorted quadrupole is combined with spiral
geodesic motion, and by solving the  geodesic equations describing
the motion of photons numerically, limits can be derived from the
whole COBE sky map. We note that unlike the open type VII$_h$
models (for example), the closed Bianchi type IX models exhibit
neither geodesic focusing nor the spiral pattern, even in the
presence of vorticity \cite{BJS}.

A special class of Bianchi type VII$_h$ models, which have a
logarithmic dynamical evolution and $\Sigma$ that is effectively
`frozen in' at late times, were studied in \cite{BJS} and it was
concluded that limits on the shear are as strong as
$|\Sigma_0|\sim 10^{-9}$ in the real Universe. In further
work \cite{BuFS}, the large-scale cosmic microwave background
anisotropies in spatially homogeneous, globally  anisotropic
cosmologies were investigated, and  improvements on previous
bounds on the total shear in the Universe were obtained by
performing a statistical analysis  to constrain the allowed
parameters of a Bianchi model of type VII$_h$ using  the
cumulative data from COBE. Consequently, very strong upper limits
on the amount of shear (and vorticity) were obtained; limits which are typically one to
two orders of magnitude higher than constraints relying entirely
on the quadrupole. Moreover, in discarding information from higher
moments, the comparison is not sensitive to the small-scale
structure present in anisotropic models that is associated  either
with the spiral pattern or with geometrical focusing  when
$\Omega<1$. This bound (\cite{BuFS}) can be improved upon by a
factor of $\sqrt{3}$ by using different statistics \cite{kogut}.

If the CMB fluctuations about isotropy are small, then so are
deviations from spatial homogeneity and isotropy \cite{almost},
provided that the Universe after last scattering can be modelled
as exactly dust (plus collision-free radiation, and may include a
cosmological constant). Model-independent limits on the present
day strengths of large-scale shear in the Universe in the
important case of an \emph{almost} isotropic CMB, which also take
into account inhomogeneities as well (assuming constraints on the
size of temporal and spatial derivatives), can then be obtained. A
general limit on the present value of $\Sigma$ can be obtained
using the quadrupole and octopole; $\Sigma_0 \leq 10{^{-4}}$
\cite{almost}. This general limit is a much weaker limit on the
shear than discussed above, which implies that the Bianchi VII$_h$
models (for example) are too special to draw conclusions from
because of their atypical evolution. However, it was shown
\cite{almost} that stronger limits on the shear \cite{BJS} can
be recovered in a (Bianchi VII$_h$) model which is Bianchi I to
first order \cite{almost}.

It is worth mentioning that since the Einstein equations are
highly non-linear, the limit as the (anisotropic) curvatures tend
to zero (in some appropriate sense) in Bianchi models does not
necessarily produce the same results as those in a Bianchi I model
with zero curvature. Although limits can often be strengthened by
considering a particular class of Bianchi cosmological models for
which exact evolution equations are known, the results are not
generic. Indeed, for very specific models of Bianchi type I and V
which isotropize, very severe limits are obtained. In general,
specific limits are obtained in each class of SH models, which
include integrated and non-integrated effects, which combine
distorted quadrupole and spiral geodesic motion, and which include
limits from numerical and statistical analyses derived from the
full COBE sky map.

\section{Analysis}

We shall investigate anisotropic cosmological models with
pressureless matter (dust) and a cosmological constant
(assumptions which are justified since recombination). Typically,
if the shear $\Sigma$ is small throughout the evolution since
recombination, the integrated effect on the CMB is not important.
However, if the shear is not small the effect can be important.

In the case of a Bianchi type I model, and exact solution (for dust and
a cosmological constant) is again possible. We find that
\begin{equation}
\label{eq1}
\Sigma^2=   \frac{\Sigma_0{^2}e^{-3t}}{ \Sigma_0{^2}e^{-3t} +
\Omega_{\Lambda_0}e^{3t} + \Omega_0 }\approx
\frac{\Sigma_0{^2}e^{-3t}}{ \Omega_{\Lambda_0}e^{3t} + \Omega_0},
\end{equation}
where $\Sigma_0$ (for example) is the value of the shear at the
present logarithmic time $t=0$, and the constants satisfy
\begin{equation}
	\Omega_0 + \Omega_{\Lambda_0} + \Sigma_0^2 =1.
\end{equation}
The approximation can be integrated exactly to give
\begin{equation}
\int^0_{t_{ls}} \Sigma dt \approx  \frac{2\Sigma_0}{3\Omega_{0}}
\left[e^{-\frac{3}{2} t_{ls}} (\Omega_{\Lambda_0}e^{3t_{ls}}+
\Omega_0)^{\frac{1}{2}} -1 \right]
	\approx \frac{2}{3}e^{-\frac{3}{2} t_{ls}} 
\frac{\Sigma_0}{\sqrt{\Omega_0}},
\end{equation}
where we will use $t_{ls} = -7$ as the logarithmic time of last 
scattering.%
\footnote{The logarithmic time $t$ is defined as the logarithm of the 
length scale: $\ell = \ell_0 e^{t-t_0}$, where we have set $t_0=0$ above. 
The logarithmic time is therefore related to the cosmic time through 
$\frac{dt}{dt_{\rm cosmic}} = H$, where $H$ is the Hubble scalar.
To determine the value of $t_{ls}$, we take the temperature of the last 
scattering surface to be $3000 K$. Using the linear relation between the 
CMB temperature and $\ell^{-1}$, we obtain
\[
	t_{ls} = \ln\left(\frac{\ell_{ls}}{\ell_0}\right)
	= -\ln\left(\frac{T_{ls}}{T_0}\right)
	= -\ln(3000/2.7) \approx -7.
\]
The value of $t_{ls}$ can be adjusted for different values of the 
temperature of the last scattering surface.
}
Imposing the quadrupole limit%
\footnote{As an idealization, we assume that the 
last scattering temperature is isotropic and that all the 
quadrupole anisotropy in the CMB comes from the integrated shear.
We use the upper bound (\ref{bound}) as a rough (order of magnitude) 
approximation of the bound on the quadrupole, $a_2 < 10^{-5}$, in order to 
forgo numerically calculating the CMB map. The integral in (\ref{bound}) 
is equal to
\[
	\int_{(t_{\rm cosmic})_{ls}}^{(t_{\rm cosmic})_0} 
\frac{\sigma}{\sqrt{3}} d t_{\rm cosmic}
\]
when expressed as an integral of the shear $\sigma$ over cosmic time.} 
\begin{equation}
\label{bound}
	\int^0_{t_{ls}} \Sigma dt < 10^{-5},
\end{equation}
we then obtain
\begin{equation}
\Sigma_0 < 4.13 \times 10{^{-10}} \sqrt{\Omega_0}.
\end{equation}
Therefore, the limit is strengthened (for a smaller value of
$\Omega_0$, corresponding to a larger $\Omega_{\Lambda_0}$) over
the case with no cosmological constant (for example, by a factor of about 2 if $\Omega_0 
= 0.3$). The physical reason for this strengthening is that for a more 
dominant cosmological constant (as measured by $\Omega_{\Lambda_0}$), 
the Hubble-normalized shear $\Sigma$ tends to zero  faster (see 
equation (\ref{eq1})). As a result, given a fixed amount of integrated 
shear (e.g., $10^{-5}$), the present day value $\Sigma_0$ is smaller in the 
presence of a dominant $\Omega_{\Lambda_0}$.

The limit on the shear in Bianchi I models from the quadrupole is
strengthened when a non-zero  cosmological constant is present.
There will be similar results for Bianchi type V models, which also 
isotropize to the future (and, in addition, for Bianchi type VII$_h$ models). 
However, for general models which do not
isotropize to the future (without the presence of dark energy),
the limits on the shear will become much much tighter.

As an illustration, let us consider the locally-rotationally
symmetric (LRS) Bianchi type III cosmological models.
The evolution equations for expansion-normalized variables can be obtained 
by setting $k=1/2$ in \cite[Section 2.1.2]{NUW} and including a 
cosmological constant:
\begin{align}
	\Sigma_\times' &= (q-2)\Sigma_\times - \Omega_k
\\
	\Omega_k' &= 2(q+\Sigma_\times)\Omega_k
\\
	\Omega_\Lambda' &= 2(q+1)\Omega_\Lambda,
\intertext{where}
	q &= 2 \Sigma_\times^2 + \tfrac12 \Omega - \Omega_\Lambda
\\
	\Omega &= 1-\Sigma_\times^2 -\Omega_k - \Omega_\Lambda.
\end{align}
Here, $\Sigma_\times$ is the shear, $\Omega_k$ is the spatial curvature, 
$\Omega_\Lambda$ is the density parameter for the cosmological constant,
$\Omega$ is the density parameter for dust, and $q$ is the deceleration 
parameter.

\begin{figure}[ht]
  \begin{center}
    \epsfig{file=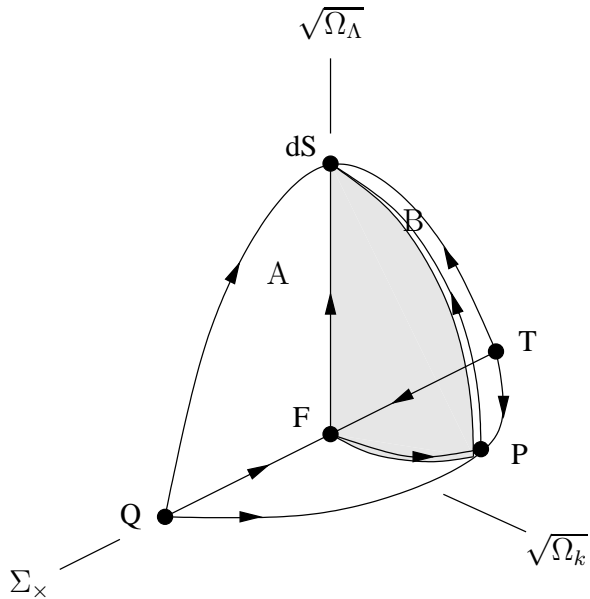}
\setlength{\unitlength}{1mm}
\begin{picture}(120,0)(0,0)
\put(20,12){\makebox(0,0)[t]{$\Sigma_\times$}}
\put(60,87){\makebox(0,0)[t]{$\sqrt{\Omega_\Lambda}$}}
\put(90,17){\makebox(0,0)[t]{$\sqrt{\Omega_k}$}}
\put(53,53){\makebox(0,0)[t]{A}}
\put(71,60){\makebox(0,0)[t]{B}}
\end{picture}
    \caption{State space of LRS Bianchi type III cosmological
models with dust and a cosmological constant. Arrows indicate the 
direction of evolution. The shaded surface indicates a set of initial 
conditions at present time yielding exactly isotropic CMB.}
    \label{fig:iso}
\end{center}
\end{figure}

Shown in Figure~\ref{fig:iso} is the state space of LRS Bianchi type III 
cosmological models with dust and a cosmological constant. The state space 
shown is a quarter of a unit sphere, with equilibrium points representing 
the flat FRW solution (F), the de Sitter solution (dS), the two 
Kasner solutions (T and Q) and the plane wave solution (P) (located at 
$\Sigma_\times = -\frac12$). The FRW solution with a cosmological constant 
is represented by the vertical line F-dS. The state space is divided into 
regions A and B (separated by the surface F-P-dS), according to whether 
the shear $\Sigma_\times$ changes sign over the course of evolution.

The observed CMB temperature $T$ is given in terms of an integral of 
the quantity $s$, 
which in this case only depends on $\Sigma_\times$ and one direction 
cosine $K_2$:
\begin{gather}
	T = T_{ls} \exp\left[ - \int_{t_{ls}}^0 (1+s)\ dt \right],
\\
	s = -(1-3K_2^2)\Sigma_\times,
\\
	K_2' = -3(1-K_2^2)K_2 \Sigma_\times.
\end{gather}
Following the calculation in \cite{Lim}, the observed CMB temperature $T$ 
is then given by
\begin{equation}
	T = T_{ls} e^{-t_{ls}} \frac{e^{2\beta_0}}{\sqrt{
	e^{6\beta_0} + (1-e^{6\beta_0}) K_2^2(0)}},
\end{equation}
where $\beta_0$ is the integrated shear:
\begin{equation}
	\beta_0 = - \int_{t_{ls}}^0 \Sigma_\times \ dt.
\end{equation}
If $\beta_0=0$, then the CMB temperature is exactly isotropic.

If the shear variable $\Sigma_\times$ does not change sign (e.g., for
initial conditions in the region B), then the viable models stay
close to FRW during the whole evolution. Numerically integrating
the null geodesic equations shows that, as expected, the limits on
the shear are much tighter in the presence of a cosmological
constant, tending to the Bianchi I constraints in the limit of
negligible curvature.

It is known that there are spatially homogeneous cosmological
models whose CMB temperature is measured to be exactly isotropic
(by all fundamental observers), and hence indistinguishable from
FRW, at one instant of time, but whose rate of expansion is highly
anisotropic \cite{Lim} (recall that we only observe the CMB at one
instant of time). These results do not contradict bounds on the
shear found by \cite{shear,BJS,Barrow}, who restrict attention ab
initio on spatially homogeneous models close to FRW. In
particular, in the Bianchi type III models under investigation but
with dust only (and no cosmological constant), if $\Sigma_\times$ can 
change 
sign (i.e., for initial conditions in the region A [very close to 
the orbit F-P]), then the models
need not stay close to FRW during the whole evolution and it is
possible for the CMB to be exactly isotropic at one instant of
time. From Figure~\ref{fig:iso}, the curve at the base of the shaded 
surface shows that $\Sigma_0$ can be very large (close to one half), while 
giving an exactly isotropic CMB. Initial conditions in a small 
neighbourhood of this curve gives a close-to-isotropic CMB.

Similarly, in Bianchi type III models with dust and a positive
cosmological constant, there is a surface of initial conditions
yielding an exactly isotropic CMB and a small neighbourhood of
this surface gives a close-to-isotropic CMB. The value of
$\Sigma_0$ on this surface is smaller than the value of $\Sigma_0$
on the curve (i.e., adding a cosmological constant tightens the
bound on $\Sigma_0$). Note that the sign change in $\Sigma_\times$
occurs (during the evolution) for all initial conditions on the
indicated surface. As a result, imposing a limit on the CMB
quadrupole will yield a looser bound on $\Sigma_0$ than in Bianchi
I models (or any model in which the shear components cannot change
sign; a sign change can lead to a cancellation in the integral
$\int \Sigma dt$).

The same argument applies to non-LRS models, and generally in
Bianchi models of all types, where there are sign changes in the
shear \cite{Lim}. However, complications in the analysis can arise
because there may be no exactly isotropic surface
 to base the calculation on. In addition, obtaining
constraints based upon the octopole may also pose problems, since
$\int \Sigma dt$ may be a poor approximation for the octopole.

\section{Discussion}

 The WMAP data provide some of the most accurate
measurements yet of the CMB and contribute to high accuracy
determinations of cosmological parameters \cite{bennett:2003}.
However, there are several studies that show that at large scales,
the CMB is not statistically isotropic and Gaussian
\cite{deOliveira-Costa:2004}. In
\cite{jaffe:2005a}, a possible detection of a correlation between
several independently discovered anomalies in the CMB sky measured
by the WMAP data and a Bianchi Type VII$_h$ template \cite{BJS}
was reported. The predictions for the CMB anisotropy patterns
arising in Bianchi type VII$_h$ universes which include a dark
energy component were presented in \cite{JH}, and it was found
that including a term $\Omega_\Lambda> 0$, can lead to the same
observed structure as in the best-fit Bianchi type VII$_h$  model
of \cite{jaffe:2005a}. Although the best-fit Bianchi type VII$_h$
model is not compatible with measured cosmological parameters, it
was shown that removing a Bianchi component from the WMAP initial
data release can account for several large angular scale anomalies
and yield a corrected sky that is statistically isotropic. It was
concluded that in the absence of an unknown systematic effect
which could explain both the anomalies and the correlation, the
WMAP data require an addition to the standard cosmological model
that resembles the Bianchi morphology \cite{jaffe:2005a}.

We have only discussed orthogonal spatially homogeneous  models here. Expanding
SH perfect fluid cosmological models with a
constant equation of state parameter $\gamma$ with a non-geodesic
`tilting' fluid congruence were studied in \cite{tilt}. It was
shown that for stiff equations of state the tilt can give rise to an 
effective quintessential, or even phantom-like, equation
of state. More importantly, it is of interest to investigate the
effect of tilt on CMB radiation observations from the perspective
of the observers moving with the fluid matter congruence and, in particular,
to determine whether the tilt can significantly affect the constraints
on the shear in SH models.

\subsection{acknowledgements} We would like to acknowledge
funding, in part, from NSERC of Canada. We thank Sigbjorn Hervik for 
discussions.

\end{document}